\documentstyle[12pt,epsf]{article}
\begin{document}
\newcommand{\preprint}[1]{\begin{table}[t]  
           \begin{flushright}               
           \begin{large}{#1}\end{large}     
           \end{flushright}                 
           \end{table}}                     
\baselineskip 22pt
\preprint{TAUP-2419-97\\hep-th9704096}

\title{Stringy Corrections to   Kaluza-Klein  Black Holes}
\author{N. Itzhaki\thanks{Email:sanny@post.tau.ac.il}
\\ \em Raymond 
and Beverly Sackler Faculty of Exact
 Sciences\\ \em School of Physics and Astronomy\\ \em Tel Aviv University,
  Ramat Aviv, 69978, Israel}

\maketitle
\newcommand{\re}{Reissner-Nordstr$\ddot{\mbox{o}}$m $\:$}
\newcommand{\tr}{\bigtriangledown}
\newcommand{\al}{{\alpha^{'}}}
\newcommand{\be}{\begin{equation}}
\newcommand{\ee}{\end{equation}}
\newcommand{\beq}{\begin{eqnarray}}
\newcommand{\eeq}{\end{eqnarray}}
\newcommand{\nn}{\nonumber}
\newcommand{\leqx}{\,\raisebox{-1.0ex}
{$\stackrel{\textstyle <}{\sim}$}\,}
\newcommand{\geqx}{\,\raisebox{-1.0ex}{$\stackrel{\textstyle >}
{\sim}$}\,}
\newcommand{\w}{Schwarzschild $\:$}
\newcommand{\th}{\theta}
\newcommand{\va}{\varphi}
\newcommand{\de}{\Delta}
\newcommand{\x}{\tilde{x}}
\newcommand{\bl}{\hspace{-.65cm}}

\begin{abstract}

We consider  string theory corrections to  4D black holes
which solve the  5D vacuum Einstein equations.
We find that the corrections vanish only for the extremal electric
solution. 
We also show that for the non-extremal electric black hole the 
 mass corrections are related to the charge corrections.
The implications  to string states counting and the correspondence
principle for black holes and strings are discussed. 

\end{abstract}
\newpage

\section{Introduction}

Although the curvature at the horizon of a large black hole is
 very small (of the order of $1/M^2$),  there are strong
gravitational interactions near the horizon due to the large
 red shift (for a recent review see \cite{tho}).
Therefore, it is generally admitted that non-perturbative
gravitational effects are involved in the full resolution,
 yet to be found, of the quantum black hole puzzle.

However, for  certain classes of black holes it seems that string
theory overcomes, at least partially, this difficulty.
In particular, for certain classes of black holes string 
states counting at small coupling yields $e^{A/4}$
\cite{ext,next}.
This result is  puzzling because Hawking-Bekenstein entropy is
obtained 
 at a region where the string size is much larger than the 
size of the 
horizon, which means that 
classically the area of the horizon, $A$, is meaningless at that region.
A classical horizon is  formed when the string 
coupling, $g$, is of order $1/M$.
An explanation  why  the counting at $g=0$  still works at
 $g\approx 1/M$ is  needed.
For extremal BPS black holes, super-symmetry non-renormalization
arguments protect the counting as one increases $g$.
Recently, non-renormalization arguments were provided for
    near-extremal  black holes \cite{mal2} in the dilute gas
 region \cite{hor3}. 
Furthermore, it has been shown that the processes of absorption
 and emission by
near-extremal D-branes configurations are not
 renormalized \cite{das}.
Still, we have not reached a full understanding of why and when
 string states counting at $g=0$ yields $e^{A/4}$.
For instance, the counting also yields $e^{A/4}$ 
 for near extremal black holes which are not in the dilute
 gas region
 \cite{next}
  and  even for extremal non-supersymetric black hole in type 
I\@ string theory \cite{dab}.
Another interesting related issue is the success  to relate
 the black hole horizon area to the string states counting, for
 some classes of non-extremal black holes 
\cite{pol}.
The relation does not predict the numerical
 coefficient, but it does yield the correct dependence on
 the mass and charges.

The aim of the present paper
 is to examine these questions from a different point of view.
Instead of increasing $g$ from $0$ toward $1/M$ and
 analyzing the microscopic dynamics,
we  begin with the classical black hole solution and 
study the $\al$ corrections to the macroscopic properties
 of the black hole.
We start in a region where  $g\approx 1$.
The dimensionless expansion
 parameter is $\lambda =\frac{\al}{M^2}$, and we decrease $g$ 
 toward $1/M$.
In 4D
\be\label{gne} G\approx g^2\al,\ee
where $G$ is the Newton constant in four dimensions. 
Therefore, as $g$ is decreased toward $\approx 1/M$ $\lambda$ is  increased toward $1$ ($G$ is fixed).
This means that in general at that region $\al$ corrections
will modify the solution completely and hence the
thermodynamical properties  of the black hole will be
changed drastically.
But this cannot be the case for all black holes  because
 for  certain classes of black holes, string states counting 
at $g=0$ gives the  area of the black hole with {\em no} $\al$
 corrections.
For these classes of black holes, $\al$
 corrections should be  suppressed.

Put differently, if for a certain black hole 
the $\al$ corrections are suppressed, then one can
decrease $g$ without changing the black hole solution.
Eventually, the size of the string will be larger than the size 
of the horizon\footnote{The one example, 
 for which the horizon does 
 not become smaller as $g$ is decreased is black hole with
 magnetic Neveu-Schwarz charge.} and
then  string states counting at $g=0$
 can be trusted.
 Since the thermo-dynamical
 properties  of the black hole are not modified
 as one changes  $g$,
  the counting should give $e^{A/4}$.

For the non-extremal black holes string states counting at $g=0$
does not yield $e^{A/4}$ so $\al $ corrections are not expected 
to be  suppressed. 
However, it has been recently shown \cite{pol} that for a wide range of
 non-extremal black holes the string states
 counting leads to the
 correct dependence on mass and charge thought it does not predict  the
 numerical coefficient. 
For these black holes we do not expect to find 
 suppression of the $\al
$ corrections but
 to find that the $\al$ corrections to
 the mass are related to the $\al$ corrections to
 the charge.

The paper is organized as follows:
In Sec. 2 we describe the 4D Kaluza-Klein
 black holes which solve the 5-dimensional
vacuum Einstein equations.
In Sec. 3 we discuss string theory corrections to the vacuum
 Einstein equations in general and in Sec. 4 we demonstrate
 their effects on \w solution. 
In Sec. 5 we analyze the corrections to the
 electric black holes. 
We present an explanation why the correspondence principle for
 black holes and string is valid for the electric  solutions.
In Sec. 6 the corrections to Pollard-Gross-Perry-Sorkin monopole
 are analyzed.
In Sec. 7 we consider the corrections to the 
extremal \re solution.

\section{The Classical Solutions}

In this section  we review the  4-dimensional black holes
 which solve
 the 5-dimensional vacuum  Einstein equations \cite{gib}.
A generalization  can be found in \cite{cve}.
The starting point is therefore
 the 5-dimensional Einstein-Hilbert action
\be S_5=\frac{1}{16\pi G_5}\int d^5x\sqrt{-g_5}R,\ee
where $G_5$ is the Newton constant in five dimensions
and $g_5$ is the five dimensional metric.
 Expanding this action  around the vacuum $ M^4\times S^1$ 
, where $S^1$ is
 a circle with radius $R$, leads to
\be S_4=\int d^4x\sqrt{-g_4}\left( \frac{R}{16\pi G}-\frac{1}{4} 
\exp (2\sqrt{3}
\kappa\sigma)F_{\mu\nu}F^{\mu\nu}+\frac{1}{2} g_4^{\mu\nu}
\partial_{\mu}\sigma\partial_{\nu}\sigma\right) ,\ee
where $G=G_5/2\pi R$, $\kappa^2=4\pi G$ and $g_4$ is the four
dimensional metric.

The relation between the 5D metric and the 4D
metric, gauge fields and scalar is
\be\label{4d5} ds^2_5=e^{4\sqrt{3}\kappa\sigma}(dx^5+2\kappa A_{\mu}dx^{\mu})^2+
e^{-2\sqrt{3}\kappa\sigma}g_{\mu\nu}dx^{\mu}dx^{\nu},\ee
where $ds_5$ is the line element in five dimensions.
The equations of motion derived from $S_4$ are invariant
 under the duality transformation 
\footnote{Quantum aspects of that duality (when $\sigma =0$) were analyzed
 in \cite{haw}. 
It was shown that the rate at which charged black holes are
 created is invariant under the Electric-Magnetic duality.
The calculations are semi-classical, still they suggest that the
duality is more then a symmetry of the classical equations of
 motion.
See also \cite{des}.}
\be
\label{dua} g_{\mu\nu} \rightarrow g_{\mu\nu},\;\;
\sigma\rightarrow -\sigma,\;\; e^{2\sqrt{3}\kappa\sigma}F_{\mu\nu}
\rightarrow \ast F_{\mu\nu}.\ee

The spherically symmetric, time-independent solutions can be 
characterized by the ADM  mass $M$, the total 
electric charge $Q$ and the total magnetic charge $P$.
The scalar charge $\Sigma$ is related to $M, Q, P$ by
\be \frac{2}{3}\Sigma=\frac{Q^2}{\Sigma +\sqrt{3}M}
+\frac{P^2}{\Sigma -\sqrt{3}M},\ee
in units where $G=1$.


The black holes solutions are the following \cite{gib}
\beq
&& e^{4\sigma/\sqrt{3}}=\frac{B}{A},\\ \nn
&& A_{\mu}dx^{\mu}=\frac{Q}{B}(r-\Sigma)dt+
P\cos \theta d\phi,\\ 
&& g_{\mu\nu}dx^{\mu}dx^{\nu}=-F/\sqrt{AB}dt^2+\sqrt{AB}/F dr^2
+\sqrt{AB}(d\theta^2+\sin ^2\theta d\phi ^2),\nn
\eeq 
where
\beq 
&&F=(r-r_+)(r-r_-),\\ \nn
&&A=(r-r_{A+})(r-r_{A-}),\\ 
&&B=(r-r_{B+})(r-r_{B-}),\nn
\eeq
and 
\beq
&& r_\pm=M\pm\sqrt{M^2+\Sigma^2-P^2-Q^2}, \\ \nn
&&r_{A\pm}=\Sigma/\sqrt{3}\pm\sqrt{\frac{2P^2\Sigma}
{\Sigma-\sqrt{3}M}},
\\ \nn
&&r_{B\pm}=-\Sigma/\sqrt{3}\pm\sqrt{\frac{2Q^2\Sigma}{\Sigma+
\sqrt{3}M}}.
\eeq
Under the duality transformation  (Eq.(\ref{dua})) the solutions
are transformed in the following  way
\be Q\leftrightarrow P,\;\;\Sigma \leftrightarrow -\Sigma .\ee

The coordinate singularities at $\theta =0$ and $\theta =\pi$ can
 be  simultaneously removed only if 
\be P=l\frac{R}{\kappa},\ee
where $l$ is an integer.
The quantization of the momentum in the $y$ direction implies that
\be Q=m \frac{2\pi\kappa}{R}.\ee
So Dirac relation
\be QP=2\pi n,\ee
is satisfied.
The minimum ADM mass  of  the electric
and magnetic charges are
\be M_q=\frac{1}{R},\;\;M_p=\frac{R}{\kappa^2}.\ee  
Therefore, from Eq.(\ref{gne}) it is clear that when 
$g\rightarrow 0$ the horizon in string units vanishes.

\section{String theory corrections}

In this section we shortly review 
 the $\al$ corrections to Einstein vacuum equations.
The low-energy effective action in string theory is given by a sum of 
classical, quantum (string loops) and non-perturbative contributions
\be S=S_c+S_l+S_{np}.\ee
We consider only $S_c$.  It is important, therefore, to find the
 region where  our approximation is valid.
 We neglect higher order corrections in $g$   and 
consider only $\al$ corrections which means
that $g\ll 1 $. 
Our dimensionless expansion parameter is
 $\frac{\al}{M^2}$, 
where $M$ is the mass of the black hole. From Eq.(\ref{gne})
 it is clear that our approximation is valid only for a large black
hole  in the region where $1/M\ll g\ll 1$ ($M^2\gg
\al\gg 1$).
To leading order in $\al $ 
\be\label{cor} S_c=\frac{1}{16\pi G_D}\int d^D
x\sqrt{-g}e^{-2\phi}\left( R+4(\tr\phi)^2+\frac{\lambda}{2}
R_{\mu\nu\rho\sigma}R^{\mu\nu\rho\sigma}\right) ,\ee
where we dropped  terms which are not relevant to our discussion
\footnote{To first order in $\al $ the full expression of $S_c$, 
 including  $B_{\mu\nu}$ terms, was calculated in \cite{tsy}.}.
$\lambda=\frac{\al}{2}, \frac{\al}{4}, 0$ for bosonic, heterotic
 and type II\@ strings , respectively.
 $G_D$ is the Newton constant in $D$ dimensions.

We are after  the $\al$ corrections in the Einstein frame, 
the frame at which the area is related to the entropy. 
The action in Einstein frame is obtained by redefining the metric by a
conformal transformation involving the dilaton
\be g_{\mu\nu}\rightarrow e^{4\phi/(D-2)}g_{\mu\nu}.\ee
The action (\ref{cor}) becomes
\be\label{cor3} S=\frac{1}{16\pi G_D}\int d^Dx\sqrt{-g}
\left( R-\frac{4}{D-2}(\tr\phi)^2+\frac{\lambda}{2}
e^{-4\phi/(D-2)}R_{\mu\nu\rho\sigma}
R^{\mu\nu\rho\sigma}\right) .\ee
The equations of motion are
\beq\label{co}\Box\phi&\!\!=\!\!&
\frac{\lambda}{4}e^{-4\phi/(D-2)}N,\\ \nn
R_{\mu\nu}&\!\!=\!\!&\lambda e^{-4\phi/(D-2)}N_{\mu\nu},\eeq
where
\beq\label{cor2} && N=
R_{\mu\nu\rho\sigma}R^{\mu\nu\rho\sigma}, \\ \nn
&& N_{\mu\nu}=R_{\mu\rho\sigma\eta}
R_{\nu}^{\rho\sigma\eta}-\frac{1}{2(D-2)}g_{\mu\nu}N\ =0.\eeq
Terms which only contribute at higher orders were eliminated
for simplicity \cite{cal2}.
The critical dimensions in string theories are $10, 26$.
Nevertheless, it is sufficient to consider only  the non-compact
 space-time.
Considering a larger $D$ will not modify the first order corrections
to the metric  of the non-compact space-time \cite{cal2}.
We  illustrate that point in the next section.

\section{\w solution}

In this section we study the $\al$ corrections to 
the 4D \w solution.
\footnote{Corrections for the \w solution at $D\geq 4$
   were considered in \cite{cal2}.
Corrections to the 4D solutions with dilaton and
 axion hairs were analyzed
in \cite{cam}.}
These corrections were  already discussed in details in
 \cite{cal2}
from the 4-dimensional point of view.
Here we consider the corrections from the 5-dimensional point of 
view.
We shall see why taking $D=5$ in Eq.(\ref{cor2}) leads
 to  the same 4D
result as taking $D=4$.

From the 4D point of view
the zeroth order  line element is 
\be ds^2=-(1-\frac{2M}{r})dt^2+\frac{1}{1-\frac{2M}{r}}dr^2+r^2
(d\theta^2+\sin ^2\theta d\phi ^2),\ee
the zeroth order value of the dilaton is $0$.

Since $N_{\mu\nu}=0$ there are no corrections at the first
order.
Note that this is a result of the  Gauss-Bonnet theorem
in 4D:
\w solution is a solution to the 4D equations, $R_{\mu\nu}=0$ and 
since in 4-dimensions one can replace the
$R_{\mu\nu\rho\sigma}R^{\mu\nu\rho\sigma}$ term in 
Eq.(\ref{cor3}) by $R^2-4
R_{\mu\nu}R^{\mu\nu}$, there are no corrections at that order.
There are corrections at higher orders which modify completely the
solution at $g\approx 1/M$ ($\al\approx M^2$).
Therefore, string states counting at $g=0$ cannot yields 
 the Bekenstein-Hawking entropy.
Namely, interactions plays an important role 
in the microscopic description of the \w black hole entropy.

From the 5D point  of view
the zeroth order   line element is 
\be\label{str} 
 ds^2=-(1-\frac{M_s}{r})dt^2+\frac{1}{1-\frac{M_s}{r}}dr^2+r^2
(d\theta^2+\sin ^2\theta d\phi ^2)+dy^2,\ee
where $y$ is  the Kaluza-Klein direction.
Eq.(\ref{str}) is a solution to $R_{\mu\nu}=0$ in 5D,
but in 5D one cannot replace the 
 $R_{\mu\nu\rho\sigma}R^{\mu\nu\rho\sigma}$ term in
 Eq.(\ref{cor3}) by a linear combination of $R^2$ and $R_{\mu\nu}R^{\mu\nu}$.
Therefore, $N_{\mu\nu}$ does not vanish
 and there are corrections.
The non-vanishing  components of  $N_{\mu\nu}$ are
\beq \label{sch} 
&& N_{tt}=Ug_{tt}, \\ \nn
&& N_{rr}=Ug_{rr}, \\ \nn
&& N_{\theta\theta}=Ug_{\theta\theta},\\ \nn
&& N_{\phi\phi}=Ug_{\phi\phi},\\ \nn
&& N_{yy}=-2Ug_{yy},
\eeq
where 
\be U=4\frac{M_s^2}{r^6}.\ee
Since in 4D there are no corrections, it is clear from 
Eq.(\ref{4d5})
that the natural ansatz for the corrected metric is
\beq\label{str2} ds^2=\frac{1}{K}\left(- (1-\frac{M_s}{r})dt^2+
\frac{1}{1-\frac{M_s}{r}}dr^2+
r^2(d\theta^2+\sin ^2\theta d\phi ^2)\right) +K^2dy^2,\eeq
where $K=1+\lambda f(r)$ .
At order $\al $ Eq.(\ref{cor2}) yields
\beq
 \ddot{f}r(r-M_s)+\dot{f}(2r-M_s)=
\frac{16M_s^2}{r^4},
\eeq
where $\dot{f}=\frac{df}{dr}$.
The solution to this equation is 
\be f(r)=\frac{4M_s^2+3rM_s+3r^2}{9r^3 M},\ee
where the boundary  condition is such that $f$ is regular at the
horizon and $f(\infty)=0$. 
So the 5D metric is  corrected in such a way
 that the 4D metric is not corrected.
The only correction is  that the radius of the compact dimension is no
longer a constant but   a decreasing function of $r$.
For the dilaton Eq.(\ref{co}) yields
\be\label{dil}  \frac{d}{dr}\left( r(r-2M)\frac{d\phi}{dr}\right) =12
\lambda\frac{M^2}{r^4}  , \ee
the solution is \cite{cal2}
\be\label{dil2} \phi  =-\lambda\left( \frac{2M}{3r^3}+\frac{1}{2r^2}
+\frac{1}{2Mr}\right)  .\ee

\section{Electric black holes}

\subsection{Extremal electric black hole}

In this subsection we  consider the solution with $Q=2M$ and
$P=0$.
The 5D line element  is
\be ds^2=-\frac{r- 4M}{r}dt^2+dr^2+r^2(d\theta^2+\sin ^2\theta d\phi ^2)
+\frac{r+4M}{r} dy^2+4\frac{M}{r}dtdy. \ee
In \cite{tsy}  it was shown that this is an exact solution in
the bosonic (as well as in the supersymmetric) string theory. 
Indeed, for that  solution we get 
\be N_{\mu\nu}=0, \ee
 so there  are no $\al $ corrections at
the first  order in $\al $. 
There is  a general argument (which does not rest on string 
theory)
 why there are no $\al$ corrections
 at any order:  
The physical meaning of the solution is a 
5D gravitational shock wave in the y-direction \cite{gib2}.
The gravitational shock wave is obtained by boosting Eq.(\ref{str})
 to infinity along the $y$ direction 
while taking $M_{S}\rightarrow 0$ and keeping their product
 (the 4D mass)  fixed.
But in the rest frame (Eq.(\ref{str})) it is clear that when
 $M_{S}\rightarrow 0$,
 any higher order corrections  vanish,  because for
 $M_{S} = 0$ the solution is flat.
Note that
 because $M_{Sc}\rightarrow 0$, the area of the black hole
vanishes.

\subsection{Non-extremal electric black hole}

To obtain the electric black hole solution 
one can  boost Eq.(\ref{str}) along the $y$ direction
and reduce to four dimensions.
The result is \cite{gib2} a black hole with mass
\be M=M_s \left( 1+\frac{v^2}{2(1-v^2)}\right),  \ee
and  electric charge 
\be\label{rel} Q=2M\frac{v}{2-v^2},\ee
 where $v$ is the velocity in the $y$ direction.

The  $\al $ corrections to the electric black hole
 solution are the
 $\al$ corrections to Eq.(\ref{str})  boosted along the $y$
 direction and reduced to 4D.
As a result there are two kinds of corrections:1- Corrections to the
dimensional reduction, these corrections appear already at the first
order in $\al $ (Eq.(\ref{str2})).
2-  Corrections to $M_s$ which  appear only at the second order in $\al$.
From the discussion in sec.4 it is clear that
 when $\al\approx M_s^2$ the solution is completely modified
 at the horizon.
So when  $\al\approx M_s^2$ the macroscopic properties of the black hole
will be changed drastically.
Thus, string state counting at small $g$ cannot yields the right
numerical coefficient.

Nevertheless, there is one macroscopic property of the 
black hole with no $\al$ corrections-
 the ratio between the mass and electric charge.
The reason for this is simple; there are no $\al$ corrections to
the  velocity boost.
 Therefore, although there are $\al$ corrections
to $M$ and $Q$   Eq.(\ref{rel}) is an exact relation between
 $M$ and $Q$.
This is why string state counting for these black hole
  does not predict the right
numerical coefficient but does give the correct
dependence on the mass and charge \cite{pol}. 

\section{Pollard-Gross-Perry-Sorkin monopole}

The PGPS monopole \cite{ggr} is an extremal magnetic black hole 
$P=2M$, $Q=0$.
The 5D metric is 
\beq \label{pop}
&& ds^2=-dt^2+V\left( dr^2+r^2d\theta ^2 +r^2\sin ^2\theta d\phi ^2
\right) 
+\frac{1}{V}\left( dy+4M(1-\cos \theta )d\phi\right) ^2,\\ \nn
&& V=1+\frac{4M}{r}. 
\eeq
The non-vanishing components of $N_{\mu\nu}$ are
\beq
\label{gps}
&&N_{tt}=-2Wg_{tt},\\ \nn
&&N_{rr}=Wg_{rr},\\ \nn
&&N_{\theta\theta}=Wg_{\theta\theta},\\ \nn
&&N_{\phi\phi}=Wg_{\phi\phi},\\ \nn
&&N_{yy}=Wg_{yy},\\ \nn
&&N_{y\phi}=Wg_{y\phi},
\eeq
where
\be W=-32\frac{M^2}{(r+4M)^6}.\ee
Notice that PGPS monopole is dual to the extremal
 electric solution, but,  unlike  the extremal electric
 solution, there are $\al$ 
corrections to the PGPS monopole which
 break the duality.

 Setting $dt=0$ in Eq.(\ref{pop}),
 the resulting 4D metric satisfies
 $R_{\mu\nu}=0$. 
 Therefore, just like in the \w solution, Gauss-Bonnet
 theorem in 4D
 implies that  the first order corrections in $\al$ 
to the 4D metric  vanish.
 The difference is that now the  4D space is the
 Euclidean space  ($r, \theta, \phi, y$) while in the
 \w solution it was the Minkowski space
 ($t, r, \theta, \phi$).
 As a result   Eq.(\ref{sch},\ref{gps}) are 
  related under $t\leftrightarrow y$.
 This implies that the ansatz for the corrected 5D metric should be 
 \beq  && ds^2=-H^{-2}dt^2+H\left( V( dr^2+r^2d\theta ^2
 +r^2\sin   ^2\theta d\phi ^2 ) \right.\\ \nn &&\left. 
+\frac{1}{V}\left( dy+4M(1-\cos \theta )d\phi\right) ^2 \right) , \eeq
 where 
 \be H=1+\lambda h(r).\ee
  To first order in $\lambda$ we get from Eq.(\ref{co}) 
 \be r\ddot{h(r)}+2\dot{h(r)}=64\frac{M^2}{(r+4M)^5}.\ee
 The boundary conditions are such that at the horizon $h$ is
regular and $h(\infty)=0$.
The solution is
 \be h(r)=-\frac{1}{12}\frac{48M^2+12rM+r^2}{M(r+4M)}.\ee
 For the dilaton Eq.(\ref{co}) yields
 \be \frac{d}{dr}\left( r^2\frac{d\phi}{dr}\right)
 =96\lambda M^2\frac{r}{(r+4M)^5} ,\ee
 The solution is
 \be\label{dil3}
  \phi=-\lambda \frac{48M^2+12rM+r^2}{8M(r+4M)^3}. \ee 
 Notice that qualitatively this solution behaves like the
 dilaton solution in \w metric (Eq.(\ref{dil2})), it
 is a negative increasing function of $r$.

 \section{The extremal \re solution}
 
 The areas of the extremal solutions which we considered 
until now  vanish.
 In this subsection we discuss the simplest extremal
 solution with  non-vanishing area-the extremal \re solution.
 This solution is self-dual at the zeroth order.
 In the previous section we saw that the $\al$ corrections 
break the  duality, so we expect the self-duality of the
 solution to be broken by the $\al$ corrections.
 The values of the zeroth order four dimensional metric,
 Abelian gauge field and scalar are
 \beq\label{sol}
 &&ds^2=-\frac{(r-r_+)(r-r_-)}{r^2}dt^2+
 \frac{r^2}{(r-r_+)(r-r_-)}dr^2+r^2d\Omega^2,
\\  \nn
&&A_{\mu}dx^{\mu}=\frac{Q}{r}dt+P\cos\theta d\phi,
\\  \nn
&& \sigma =0,
\eeq 
where  
\be\label{sll} r_{\pm}=M\pm\sqrt{M^2-Q^2-P^2},\ee
 and
\be P=Q=\frac{M}{\sqrt{2}}.\ee
The calculation of the first order corrections of this solution
relegated to the appendix.
The result is a solution which is similar to
Eqs.(\ref{sol}, \ref{sll}) but with
  
\beq\label{sol2}
&& M_g(r)=M-\lambda (r-m)\left[ \frac{1}{70r^5M}\left( 
182r^2(r-M)^2\log (\frac{r}{r-M})
\right.\right.\\ \nn &&\left.\left. 
-82M^4-22M^3r+269M^2r^2-214Mr^3-32r^4\right)-\frac{M^2(r-M)}{4r^5}\right]
,\\ \nn
&& M_i(r)=M-\lambda (r-m)\left[ \frac{1}{70r^5M}\left( 
182r^2(r-M)^2\log (\frac{r}{r-M})
\right.\right.\\ \nn &&\left.\left.
-82M^4-22M^3r +269M^2r^2-214Mr^3-32r^4\right)+\frac{M^2(r-M)}{4r^5}
\right],
 \\ \nn
&& Q(r)=\frac{M}{\sqrt{2}}-\lambda 
\frac{\sqrt{2}(r-m)}{140r^4M^2}\left[364 \log (\frac{r}{r-M})(
r^2(r-M)^2) \right. \\ \nn
&&\left. -122M^4-142M^3r+545M^2r^2-421Mr^3
\right],   \\ \nn 
&& P(r)=\frac{M}{\sqrt{2}}, \\ \nn
&& \sigma (r)=\lambda\frac{1}{105\sqrt{3}r^4M^2}
\left[ 1092r^2(r-M)^2\log  (\frac{r}{r-M})+1204r^3M-1697r^2M^2
 \right.\\ \nn &&\left.
+360rM^3+225M^4\right]  .
 \\ \nn
\eeq
where $M_g$ and $M_i$ are the gravitational and inertial mass
 which are defined respectively as
\be -g_{tt} =1-\frac{2M_g}{r}+O\left( \frac{1}{r^2}\right),\;\;
g_{rr}=1+\frac{2M_i}{r}+O\left( \frac{1}{r^2}\right) .\ee
The boundary conditions  that we choose are
\be\label{bon}
 M_g=M_i=\sqrt{2}P=\sqrt{2}Q=M.\ee
It is convenient to work with these boundary conditions
because the  temperature remains
zero so the black hole remains extremal.
At infinity one finds  the  corrections to the total mass and 
charges 
\beq
&& M_i(\infty )=M_g(\infty )=M+\lambda\frac{16}{35M}, \\ \nn
&& Q(\infty )=\frac{M}{\sqrt{2}}+\lambda\sqrt{2}\frac{57}{140M},
 \\ \nn 
&& P(\infty )=\frac{M}{\sqrt{2}}, \\ \nn
&& \sigma (\infty )=0. \\ \nn
\eeq
Note that due to 
 the equivalence principle  one gets
 $M_g(\infty)=M_i(\infty)$. 
We see, therefore, that there are $\al$ corrections to the black
 hole extremality condition, which become of order one when
 $g\approx 1/M$ $(\lambda\approx M^2$).
Thus for this black hole string states 
counting at small $g$ cannot yields 
 the Bekenstein-Hawking entropy.
 Moreover, 
\be M(\infty)\neq \sqrt{2}Q(\infty),\;\;M(\infty)\neq
\sqrt{2}P(\infty),\;\;Q(\infty)\neq P(\infty),\ee
so the correspondence principle for black holes and strings 
\cite{pol}
is not expected to be valid for this black hole.
Since $\al $ corrections break the duality it is unlikely that 
the correspondence principle for black holes and strings will 
be valid for $p\neq 0$.

For the dilaton Eq.(\ref{co}) yields
\be \frac{d}{dr}\left( (r-m)^2\frac{d\phi}{dr}\right)=3\lambda M^2
\frac{8r^2-16rM+7M^2}{2r^6}, \ee
the solution is 
\be \phi =\frac{\lambda}{40M^2r^4}\left( 60r^4\log
  (\frac{r}{r-M})-64r^3M-34r^2M^2-24rM^3+21M^4\right ). \ee
At $r\gg M$ this solution, like the corrected dilatons in
 \w and GPS solutions
(Eq.(\ref{dil2},\ref{dil3})),  is negative  so $\phi $ decreases
as $r$ is decreased.
But, unlike Eq.(\ref{dil2}, \ref{dil3}) here at the horizon 
\be
 \lim_{r\rightarrow M^+} \phi=\infty ,\;\;\;
 \lim_{r\rightarrow M^-} \phi=-\infty. \ee
The physical meaning of that solution 
 can be understood from Eq.(\ref{cor3}). 
Just outside the horizon the effective string scale
$\al e^{-\phi /2}$ goes to zero
 , while just inside the horizon $\al e^{-\phi /2}\rightarrow\infty$ .
It is worthwhile to mention that for the non-extremal \re
 solution the
corrected dilaton is not singular at the horizon.    

\section{Summary }

The black holes that we analyzed here are rather simple,
 from string
theory point of view, because at the zeroth order  
$B_{\mu\nu}=0$ and $\phi =0$ .
As a result, the only solution with no $\al $ corrections is the
extremal electric black hole \cite{tsy}.
At its neighborhood the $\al $ corrections are suppressed.
All other solutions, even the extremal solutions,  are
 completely modify when $g\approx 1/M$, thus
we do not expect that the microscopic description of these
 black holes at $g=0$ will yield $A/4$.
It would be interesting to consider the $\al $ corrections
to   solutions with
$B_{\mu\nu}\neq 0$ and $\phi\neq 0$ at the zeroth order.
 Of special interest are the near-extremal black holes 
 for which the string states counting at $g=0$ gives $A/4$.
For these solutions $\al$ corrections should be suppressed.

Another conclusion is that the correspondence principle for
 black holes and strings is not valid for all  black holes
 solutions.
In fact, our analysis of the $\al $ corrections for
the  Kaluza-Klein black holes implies that 
 it is valid only for pure electric black holes.
It would be interesting to  compute the  $\al $ corrections
  to other  black holes  for which the correspondence 
principle is valid \cite{pol}.
It might provide us with an indication about the crucial
 properties of black holes for which the correspondence
 principle of
 \cite{pol} is valid. 

It is important to emphasis that our conclusions are valid 
only for  bosonic and heterotic string theories.
In type II\@ theories there are corrections only at order 
$\al ^3$ \cite{gro1}.
We did not perform the full calculations in that case but we 
expect to get the same qualitative result as the following
 argument implies.
$\al$ corrections to Einstein equations in  type II\@ theories
 in the string frame
lead in out case to
$R_{\mu\nu}=\al ^3\tr_{\mu}\tr_{\nu}F(r)$. 
This term cannot be canceled by the 
dilaton contribution which  is proportional to  $g_{\mu\nu}$.
So locally $\al $ corrections are not suppressed.
Still, there might be a global effect which would cancel the 
corrections at infinity to the ADM mass and charges.

\vspace{1cm}

\bl {\bf Acknowledgments } I would like to thank A. Casher
 and S. Yankielowicz for helpful    discussions.  
\newpage
\appendix
\section{Appendix}
In this section we derive Eq.(\ref{sol2}).
There are many ways one can calculate the first order corrections
 to Eq.(\ref{sol}, \ref{sll}).
We choose to work with the 4D variables $M_g, M_i, Q$ and $P$.
To the first order in $\al $ the component $(r,\theta)$ of Eq.(\ref{cor2}) yield
\be \dot{P}=0,\ee
where $\dot{f}=\frac{df}{dr}.$
From Eq.(\ref{bon}) we get
\be P=\frac{M}{\sqrt{2}}.\ee
From the equivalence principle it is clear that one should get
$M_g(\infty )= M_i(\infty )$, therefore it seems useful to work with
\be M_+ =\frac{1}{2}(M_g +M_i),\;\;M_- =\frac{1}{2}(M_g -M_i),
\ee 
Instead of $M_g$ and $M_i$.
Gauss law implies that it might be   helpful to work with 
\be e(r)=\frac{(Q(r)-M/\sqrt{2})}{r},\ee
and not with $Q(r)$.
We also work with
\be k(r)=-\frac{2}{\sqrt{3}}\sigma(r)\ee
The nontrivial independent components of Eq.(\ref{cor2}) to $\lambda$ 
order are
\beq\label{soll}
(tt):\;&&
\ddot{M_+}\frac{(r-M)^2}{r^3}+\frac{\sqrt{2}M}
{2r^4}\left( 3r\ddot{e}(r-M)^2+4\dot{e}(r-M)^2\right)
\\ \nn &&
+\frac{1}{2r^6}\left( \ddot{k}r^2(r^2-2rM+5M^2)(r-M)^2
\right. \\ \nn && \left.
+2\dot{k} r(r-M)
(r^3-2r^2M-rM^2+6M^3)+6M^2k(M^2+r^2)\right) 
\\ \nn &&
+\frac{1}{r^5(r-M)^2}\left(
  \ddot{M_-}r^2(r-M)^4-2rM\dot{M_-}(r+M)(r-M)^2
\right. \\ \nn && \left.
+2M M_-(Mr^2+3M^3-M^2r+r^3)\right) 
\\ \nn &&
=M^2\frac{2r^4-36r^3M+97r^2M^2-94rM^3+37M^4}{r^{10}} 
\\ \nn
(t\phi):\;\;\; &&
\frac{\sqrt{2}M}{r^3}\left(\ddot{e}r(r-m)^2+2\dot{e}(r^2-2rM-M^2)\right)
\\ \nn &&
+\frac{M^2}{r^4}\left(2r\ddot{k}(r-M)^2+\dot{k}(r^2+2rM-3M^2)\right )
\\ \nn &&
-2\frac{M^2}{r^4(r-M)^2}\left( r\dot{M_-}(r-M)^2-M_-(3M^2+r^2)\right)
\\ \nn &&   
 =-2M^3\frac{6r^3-14r^2M+10rM^2-5M^3}{r^9}
\\ \nn 
(rr):\;&&
-r\ddot{M_+}+\frac{M}{\sqrt{2}}(r\ddot{e}+4\dot{e})
-\frac{1}{2r^2}\left(
  \ddot{k}r^2(r-M)^2+2\dot{k}r^2(r-M)+6M^2k\right)
\\ \nn &&
-\frac{1}{r(r-M)^2}\left(r^2\ddot{M_-}(r-M)^2+
2r\dot{M_-}(3rM-M^2-2r^2)
\right. \\ \nn &&\left.
+2M_- (2r^2+rM-M^2)\right) 
\\ \nn &&
=M^2\frac{2r^2-3M^2}{r^6} 
\\ \nn 
(\theta\theta):\; &&
-2\dot{M_+}+\sqrt{2}M\dot{e}-\frac{1}{2r^2}\left(
  \ddot{k}r^2(r-M)^2+2\dot{k}r^2(r-M)-6M^2k\right)+
\\ \nn &&\frac{2(r+M)}{r(r-M)}M_-
=M^2\frac{8r^2-20rM+11M^2}{r^6}
\\ \nn 
(yy):\; &&
-\sqrt{8}M\dot{e}+\ddot{k}(r-M)^2+2\dot{k}(r-M)
+4M_-\frac{M^2}{r(r-M)^2}=
\\ \nn &&
-M^2
\frac{5M^2-16rM+8r^2}{r^6}  
\eeq
Note that due to the Bianchi identities we have five equations but
 only four variables ($M_{\pm}, e, k$).
To solve  this equations it is helpful to notice that the combination
\be(t\phi)-\frac{M^2}{r^3}(yy)-\frac{r}{2}\left(
  (tt)+\frac{(r-M)^2}{r^4}(rr)\right), \ee 
gives the simple differential equation
\be
\frac{2}{r^4}\left(r \dot{M_-}(r-M)-M_-(r+M)\right)
=2\frac{M^2}{r^9}(r-M)^3
.\ee
whose solution is
\be
M_-=\frac{M^2(r-M)^2}{4r^5}
\ee
Plugging this solution  back into  Eq.(\ref{soll}) 
one gets equations for $M_+, e, k$ whose solution is
 Eq.(\ref{sol2}).

\end{document}